\documentclass[twocolumn,aps,showpacs,amssymb,superscriptaddress,10pt]{revtex4}
\usepackage{graphicx}
\usepackage{subfigure}
\usepackage{amsmath,amssymb,ascmac,color}
\usepackage{type1cm}
\usepackage{bm}

\newcommand\beq{\begin{equation}}
\newcommand\eeq{\end{equation}}
\newcommand\beqa{\begin{eqnarray}}
\newcommand\eeqa{\end{eqnarray}}

\newcommand{\ds}[1]{#1 \hspace{-0.5em}/}  
\newcommand{\Ds}[1]{#1 \hspace{-0.55em}/} 
\newcommand{\Dds}[1]{#1 \hspace{-0.75em}/} 

\newcommand\bp{{\bf p}}
\newcommand\bk{{\bf k}}
\newcommand\br{{\bf r}}
\newcommand\bq{{\bf q}}
\newcommand\bx{{\bf x}}

\begin{document}

\title{Quark beta decay in the inhomogeneous chiral phase and cooling of compact stars}

\author{T. Tatsumi}
\affiliation{Department of Physics, Kyoto University, Kyoto 606-8502, Japan}

\author{T. Muto}
\affiliation{Department of Physics, Chiba Institute of Technology,2-1-1 Shibazono, Narashino, Chiba 275-0023, Japan}

\begin{abstract}
A novel cooling mechanism is proposed for neutron stars,  
based on the recent development in the studies of the QCD phase diagram. 
Possible appearance of the inhomogeneous chiral phase makes the quark beta decay 
without gluonic interaction. An estimate of the neutrino emissivity shows the order of 
$10^{24-26}(T/10^9)^6$(erg$\cdot$cm$^{-3}$ $\cdot$ s$^{-1}$) near the phase boundaries, 
whose efficiency is comparable with the usual quark cooling or pion cooling, but it works only 
in the limited density region. These features may give another cooling scenario of neutron stars.
 
\end{abstract}


\maketitle

\section{Introduction}

The appearance of the inhomogeneous phases near the phase boundary should be rather common phenomenon in many-body systems. Fulde-Ferrell-Larkin-Ovchinikov (FFLO) state is one of the typical examples in superconductivity in the presence of magnetic impurities 
\cite{super}
and has been recently studied in dilute atomic gas \cite{atom}, or within the context of color superconductivity in QCD \cite{raj}. 
Inhomogeneous phase formation in magnetic materials is another one; spin density wave \cite{sdw,grus} or texture \cite{mag}. 
Similar subject has been also addressed in the QCD phase diagram. The deconfinement and chiral transition have been studied both 
theoretically and experimentally in the QCD phase diagram \cite{fuk}. The direct numerical calculation based on the lattice QCD theory should be  
a most powerful tool for this purpose, but its validity is, for the present, limited to high temperature and low density 
region due to the sign problem. On the contrary, the phase structure is also important and interesting in the high-density 
region in the light of recent progress in the observation of compact stars \cite{sch}. 
Many theoretical studies have been devoted to the chiral transition by the use of the effective models of QCD \cite{fuk}. 
Consequently, spontaneous symmetry breaking (SSB) should be restored at high density, which is specified by the vanishment of the $q\bar q$ scalar condensate, $\langle{\bar\psi}\psi\rangle$: it is the order parameter in the chiral transition and takes a finite value in the vacuum to generate the quark or nucleon mass. In these studies it is implicitly assumed that the condensate is scalar and uniform, while Lorenz invariance or parity symmetry no more holds at finite densities.         

Recently there appeared many papers about the possibility of the inhomogeneous chiral phases \cite{chi}, where the condensates are not restricted to the scalar one and they are spatially nonuniform, stimulated by the mathematical discoveries of the Hartree-Fock solutions in the 1+1 dimensions \cite{bas};
it has been shown that analytic solutions are obtained in terms of the elliptic functions in the Gross-Neveu model or two dimensional NJL model
 in the large $N$ limit. The order parameter or the mean-field is generalized to be complex as 
$M(x)=\langle\bar\psi \psi\rangle+i\langle\bar\psi i\gamma_5 \psi\rangle=\Delta(x)e^{i\theta(x)}$, they have found the solutions of the self-consistent coupled-equations of quark and $M(x)$ for these models. 
Its direct application is possible for the one dimensional order in 1+3 dimensions by embedding the one dimensional structure and operating the Lorentz boost in the direction perpendicular to it. Actually Nickel have performed this procedure for the real kink crystal (RKC) \cite{nick}, where $\theta(\br)=0$. Similar procedure may be also possible for the chiral spiral. The chiral spiral has a former history. 
Nakano and one of the authors (TT) have studied the possibility of the inhomogeneous chiral phase in 1+3 dimensional quark matter within the $SU(2)\times SU(2)$ NJL model \cite{dcdw}. Using $\theta(\br)=\bq\cdot\br$, the chiral condensates take form, $\langle\bar\psi \psi\rangle=\Delta\cos(\bq\cdot\br),
\langle\bar\psi i\gamma_5\tau_3 \psi\rangle=\Delta\sin(\bq\cdot\br)$, which is a 1+3 dimensional realization of the chiral spiral in 1+1 dimensions. They called it dual-chiral-density wave (DCDW). 
Since the spatial displacement of the condensates is compensated by chiral rotation on the quark field, the external degrees of freedom is mixed with the internal ones;  the wavefunction changes  
$\psi\rightarrow e^{i\bk\cdot {\bf d}}{\rm exp}(i\gamma_5\tau_3 \bq\cdot{\bf d}/2)\psi$ following the displacement, $\br\rightarrow \br+{\bf d}$. 

The physical mechanism has been discussed in ref.\cite{dcdw}; the nesting effect of the Fermi surface may play a key role as in condensed matter physics \cite{sdw,grus,cdw,gruc,grud}. If this is the case, the appearance of the inhomogeneous phase should be rather robust and less model-dependent. However, there are still left many subjects to be elucidated.
In ref.\cite{nick} Nickel suggested that RKC is more favorite than DCDW in symmetric quark matter in the chiral limit by comparing the thermodynamic potential. However, it should be an ideal situation and we must carefully compare both cases in realistic situations, including the model dependence \cite{car,mul}. In particular,  the effect of the quark current mass \cite{mae,kar} and magnetic field should be important \cite{fro,tat2}. Actually chiral anomaly plays an important role and DCDW develops in a wide region in the presence of the magnetic field \cite{fro,tat2}
Asymmetric quark matter or chemical equilibrium is also important in compact stars \cite{ebe}. Thus more elaborate studies are needed to say definite things about 
the most plausible configuration, the critical density or the critical temperature.

On the other hand it should be important to consider their phenomenological implications. Since the order parameter is spatially nonuniform and takes a periodic function, one may expect elasticity like a Coulomb lattice or liquid crystal \cite{gen}. The periodicity of the order parameter may give rise to another effect. The 
quark wave function accordingly takes a special form dictated by the generalized Bloch theorem \cite{bas}; momentum is not a good quantum number, so that the condensates should modify the momentum conservation in the elementary processes like the Umklapp process in solid \cite{kit}. Moreover, the appearance of the pseudoscalar condensate is related to magnetic properties \cite{dcdw,tat2}. Thus it should be interesting and important to figure out how such features manifest by confronting them with physical phenomena. 
In the relativistic heavy-ion collisions the formation of quark-gluon plasma has been expected. Some implication of the chiral critical point has been studied theoretically and experimentally \cite{fuk}. If the inhomogeneous phases are realized during the collisions, they might give rise to some phenomena never discussed yet \cite{bas2}.
In this paper we consider the cooling process in compact stars as an astrophysical implication of the inhomogeneous phases.

Cooling of compact stars has provided us with information about form of matter at high-densities \cite{bec}. 
Recent observations of the surface temperature of young pulsars have suggested that  
some compact stars such as 3C58 or Vela seem to have rather low temperature which should  
be barely explained by the standard scenario. Such stars might require exotic cooling; 
quark cooling is one of the fast cooling mechanisms in the core region. On the other hand,  
Cas A also presents important information about the thermal evolution of young pulsars \cite{casa}. 
Considering the young age of $t=330$yr,  
the observed effective temperature of Cas A also gives a strong constraint on the equation of state  
and cooling processes.
In the recent paper we have presented models which satisfy both cases of Cas A and other cooler stars 
by considering the quark matter in the core \cite{nod}. 
 
As a cooling mechanism in quark matter, the neutrino emission by way of the direct Urca process 
is well-known and standard, $d\rightarrow u+e^-+{\bar\nu}_e$
, $u+e^-\rightarrow d+\nu_e$
\footnote{If there develops color superconductivity, the phase space is exponentially restricted due to the pairing gap, so that neutrino emission is inefficient \cite{sch}. }
. 
This process works 
for interacting quarks, while it is strongly prohibited for free and light quarks 
due to the kinematical condition ({\it triangular condition}) at low temperature. The neutrino emissivity is then 
efficient and proportional to $\alpha_s T^6$ with the QCD coupling constant $\alpha_s$\cite{iwa80}. 

Here we discuss a new cooling mechanism, based on the recent development in 
understanding of the QCD phase diagram: possible appearance of the inhomogeneous  
phases near the chiral transition \cite{fuk}.  
Accordingly the chiral condensates modify the elementary process by supplying the 
extra momentum at the interaction vertex
\footnote{Rough idea has been already presented in ref.\cite{nic}.}.  

This paper is organized as follows. In Sec.~\ref{sec:framework} we present our framework for calculating the neutrino emissivity, where some characteristic aspects associated with  the DCDW phase are pointed out. In Sec.~\ref{sec:phase} numerical results for the neutrino emissivity are demonstrated, and their implications for cooling of hybrid stars are briefly discussed in Sec.~\ref{sec:discussion}. 
Summary and concluding remarks are given in Sec.~\ref{sec:summary}. Properties of the quark propagator is summarized in Appendix~\ref{sec:a}. 
The evaluation of the weak matrix element is presented in Appendix~\ref{sec:b}, and details of angular integrals for obtaining the emissivity in two limiting cases are given in Appendices~\ref{sec:c} and \ref{sec:d}. 

\section{Framework}
\label{sec:framework}

\subsection{DCDW}

\begin{figure}[htbp]
 \begin{center}
  \subfigure{\includegraphics[height=0.45\linewidth]{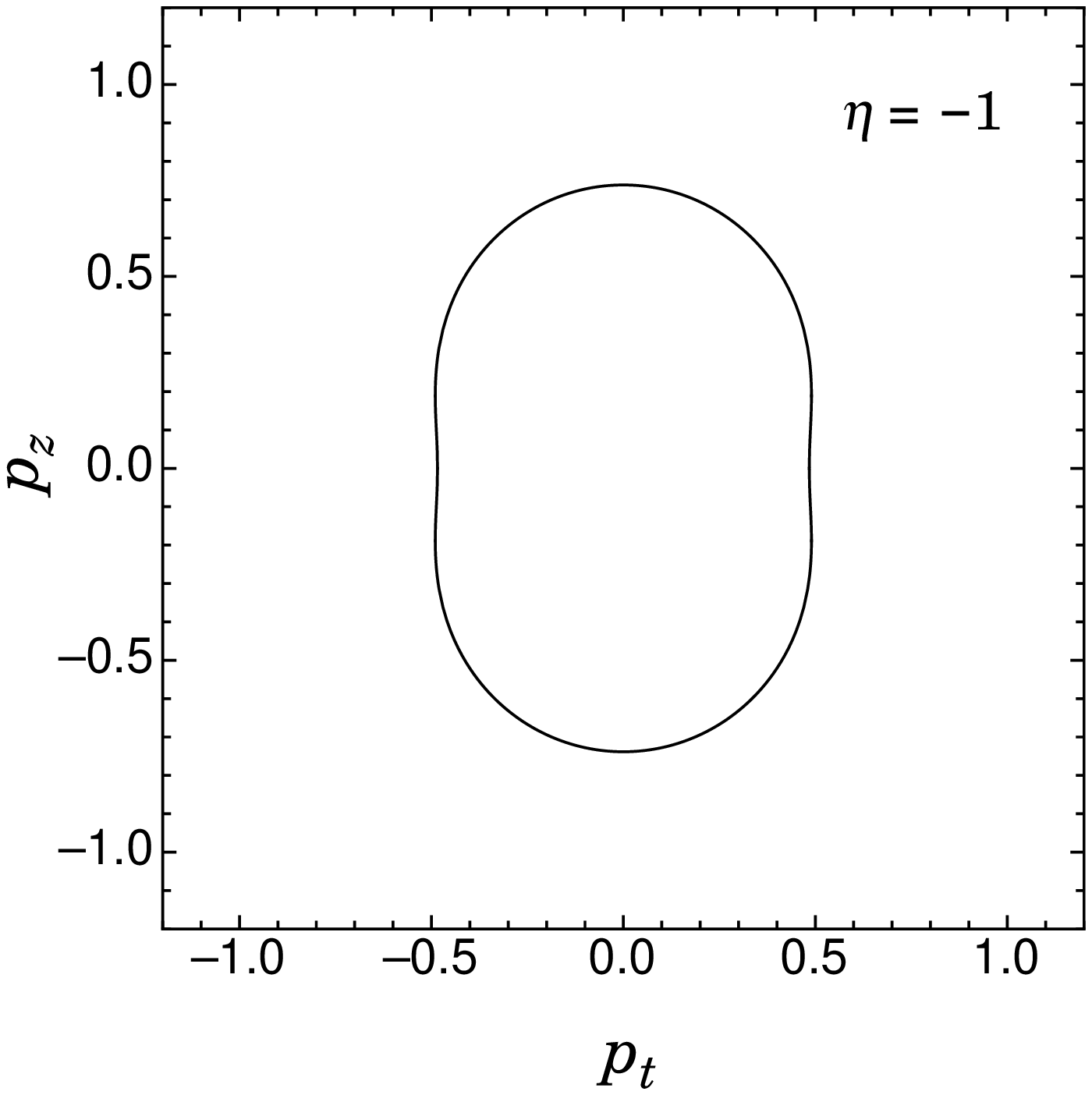}}
  \hspace{15mm}
  \subfigure{\includegraphics[height=0.45\linewidth]{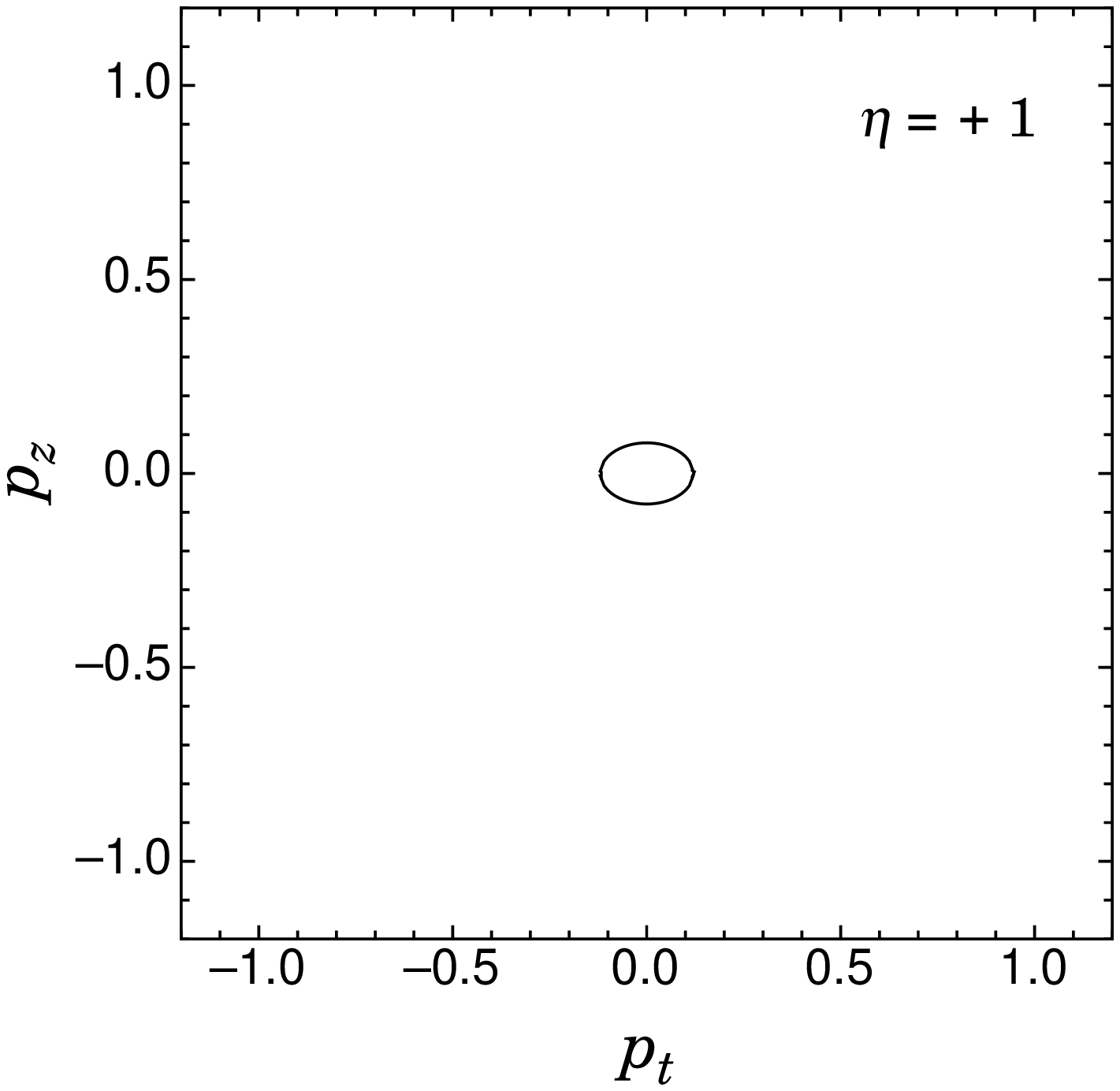}}
  \caption{Fermi surfaces at the onset density with arbitrary scale. The top panel denotes that of majority particle with the spin polarization $\eta=-1$, while the bottom panel denotes that of  minority particle with $\eta=+1$. $p_t=(p_x^2+p_y^2)^{1/2}$ and all the values of momenta are written in the unit of the cut-off parameter $\Lambda$.}
 \label{fig1}
\end{center}
 \end{figure}

First we briefly summarize the results about DCDW in the previous work \cite{dcdw}.
The DCDW phase can be represented as a chirally rotated state from normal quark matter,
\begin{eqnarray}
\left.\right|{\rm DCDW}\rangle&=&{\rm exp}\left(i\int\theta(\br) A_3^0(\br)d^3r\right)\left.\right|{\rm normal}\rangle\nonumber\\
&\equiv &U_{\rm DCDW}(\theta)\left.\right|{\rm normal}\rangle,
\end{eqnarray}
where $A_i^\mu$ denotes the axial-vector current with $i$-th isospin component. We restrict the chiral transformation to $U_{I_3}(1)$ 
around the third axis in the isospin space to preserve electromagnetic charge of the system.
Then we can easily check the following relations:
\beqa
\langle {\rm DCDW}|{\bar\psi} \psi|{\rm DCDW}\rangle&=&\Delta\cos(\bq\cdot\br),\nonumber\\
\langle {\rm DCDW}|{\bar\psi}i\gamma_5\tau_3 \psi|{\rm DCDW}\rangle&=&\Delta\sin(\bq\cdot\br),
\eeqa
for $\theta(\br)=\bq\cdot\br$, where the amplitude $\Delta$ is given by $\langle{\rm normal}|{\bar\psi}\psi|{\rm normal}\rangle$. 
In the following we use the NJL model with $SU(2)\times SU(2)$ symmetry in the chiral limit, as an effective model of QCD.
When we define the new quark field $\psi_W$ by way of the Weinberg transformation such that 
\beq
\psi_W={\rm exp}(i\gamma_5\tau_3 \bq\cdot\br/2)\psi,
\eeq
$\psi_W$ satisfies the following Hartree equation,
\beq
\left(i\ds{\partial}-m+1/2\gamma_5\tau_3\ds{q}\right)\psi_W=0,
\label{hartree}
\eeq
with $q=(0,\bq)$. Here $m=-2G\Delta$ is the dynamical quark mass generated by the quark-quark interaction with the coupling constant $G$. 
The quark eigenstate (quasi-particle) then can be represented by $|\bp,\eta, \epsilon\rangle$ with quantum numbers momentum $\bp$, $\eta=\pm 1$ specified by the spin polarization and $\epsilon=\pm 1$ the particle and anti-particle. Accordingly the energy eigenvalues read 
\beq
E^\eta_\bp=\epsilon\left(E_p^2+|\bq|^2/4+\eta\sqrt{(\bp\cdot\bq)^2+m^2|\bq|^2}\right)^{1/2},
\label{eq:energy}
\eeq
with $E_p=(|\bp|^2+m^2)^{1/2}$. Thus the Fermi surfaces of the quasi-particles are deformed in this case: one has the prolate shape and the other the oblate shape (see Figs.\ref{fig1} and \ref{fig2}).

Choosing $\bq//{\hat z}$ without loss of generality, the eigenfunction renders \cite{dcdw}
\beq
\langle \br|\bp,\eta,\epsilon=1\rangle=u^\eta_W(\bp){\rm exp}(i\bp\cdot\br)
\eeq
with the spinor
\[
u^\eta_W=
\left(
\begin{array}{l}
a_1^\eta\phi_+ +a_2^\eta\phi_- \\
b_1^\eta\phi_+ +b_2^\eta\phi_-
\end{array}
\right),
\]
where $\phi_\pm$ is the Pauli spinor s.t. $\sigma_z\phi_\pm=\pm\phi_\pm$ 
and the coefficients $a_i^\eta,b_i^\eta$ are given by
\beqa
\frac{a_1^\eta}{a_2^\eta}&=&\frac{p_-}{p_z}\cdot
\frac{m+\eta\beta}{ E_{\bp}^\eta-|\bq|/2-\eta\beta},\\
\frac{b_1^\eta}{a_2^\eta}&=&\frac{p_-}{E_{\bp}^\eta-|\bq|/2-\eta\beta},\\
\frac{b_2^\eta}{a_2^\eta}&=&\frac{p_z}{-m+\eta\beta},
\eeqa
for $\tau_3=1$ and $q_z=|\bq|$, with $\beta\equiv (p_z^2+m^{2})^{1/2}$ and $p_-\equiv p_1-ip_2$ 
for the positive-energy solutions. The negative-energy solutions ($\epsilon=-1$) 
are obtained by replacing $E_{\bp}^\eta$ by $-E_{\bp}^\eta$. 
Note that these eigenfunctions are written in terms of the newly-defined quark field $\psi_W$. 

DCDW develops between the onset chemical potential $\mu_{c1}$ and the termination one $\mu_{c2}$. 
Their values and those of the parameters are listed in Table~\ref{para}~\cite{dcdw}.

\begin{table}[htbp]
\begin{center}
\begin{ruledtabular}
 \begin{tabular}{ccccccc}
  $\mu_{c1}$ & $\mu_{c2}$ & $m_{c1}$ & $m_{c2}$ & $|\bq|_{c1}$  & $|\bq|_{c2}$    \\
  .49       & .53      & .2    & .01 & .55 & .8       \\
 \end{tabular}
\end{ruledtabular}
 \caption{Values of the chemical potentials and the parameters in the unit of the cut-off parameter $\Lambda=850$MeV. }
 \label{para}
\end{center}
\end{table}

\subsection{Neutrino emissivity}

We consider the neutrino emissivity in the presence of DCDW, following refs.\cite{max},\cite{mut},\cite{tat}.
Consider the beta decay of $d$ quarks s.t. $d(p_1)\rightarrow u(p_2)+e^-(p_3)+\bar\nu_e(p_4)$ in the DCDW phase, where $p_i=(E_i,\bp_i)$ denotes 
the four-momentum.
Taking the effective interaction as the current-current form, $H_W=\frac{{\tilde G}_F}{\sqrt{2}}h_{1+i2}^\mu l_\mu+{h.c.}$, 
the transition matrix element is given as $W_{fi}\equiv\langle u,e^-,{\bar\nu}_e\left|H_W\right|d\rangle=\langle u_W,e^-,{\bar\nu}_e\left|{\tilde H}_W\right|d_W\rangle $, where 
\beq
{\tilde H}_W=U_{\rm DCDW}(\bq)H_WU_{\rm DCDW}^\dagger(\bq)=\frac{{\tilde G}_F}{\sqrt{2}}{\tilde h}_{1+i2}^\mu l_\mu+{h.c.}. 
\eeq
${\tilde G}_F=G_F\cos\theta_C$ with $G_F$ being the Fermi weak coupling constant and $\theta_C$ the Cabibbo angle.
Here it is to be noted that the matrix element between eigenstates $|\bp,\eta,\epsilon\rangle$ should be calculated by using 
the untransformed states, ${\rm exp}(-i\gamma_5\tau_3\bq\cdot\br/2)|\bp,\eta,\epsilon\rangle$, as in the pion cooling \cite{max}.
The transformed quark current ${\tilde h}_{1+i2}^\mu$ now reads,
\beq
{\tilde h}_{1+i2}^\mu=U_{\rm DCDW}(\bq)h_{1+i2}^\mu U_{\rm DCDW}^\dagger(\bq)={\rm exp}\left(-i\bq\cdot\br\right)h_{1+i2}^\mu,
\eeq
by way of the current algebra, which implies that DCDW modifies the momentum conservation at the weak-interaction vertex
\footnote{This may remind us of the pion cooling \cite{max}. 
The pion condensed state can be generated by a combination of global chiral rotation and local isospin rotation, 
but the pion momentum is supplied at the weak-interaction vertex in a similar way.}
.
Usual triangular condition among $\bp_1,\bp_2$ and $\bp_3$ is now relaxed by the momentum supply from DCDW, 
so that the beta decay process becomes possible.

Since quarks should be treated as quasiparticles in our case, 
naive application of the emissivity based on the Fermi golden rule is not relevant: we must properly 
take into account the wave-function renormalization besides the deformation of the Fermi surface.
Thus we start with more general formula. The neutrino emissivity can be then given as \cite{kad,jai}
\beqa
\epsilon&=&N_c{\tilde G}_F^2\int\frac{d^3 p_3}{(2\pi)^3 2E_3}\int\frac{d^3p_4}{(2\pi)^3 2E_4}E_4L_{\lambda\sigma}
\nonumber\\
&\times& n_F(-E_3+\mu_e)f_B(k_0){\rm Im}\Pi_R^{\lambda\sigma}(k),
\eeqa
with $k=(E_3+E_4-\mu_e,\bp_3+\bp_4)$, the Fermi-Dirac distribution function $n_F$ and the Bose-Einstein one $f_B$.
The information of the quark tensor is summarized in the $W$ boson polarization tensor,
\beq
\Pi_R^{\lambda\sigma}(k)=T\sum_n\int\frac{d^3p_1}{(2\pi)^3}{\rm tr}\left[\Gamma^\lambda S_W^d(p_1)\Gamma^\sigma S_W^u(p_1-k\pm q)\right]
\eeq
with $\Gamma^\mu=\gamma^\mu(1-\gamma_5)$ and the quark propagator, $S^{-1}_W(p)=\ds{p}-m+\gamma_5\tau_3\ds{q}/2$. Since the contribution from the Dirac sea  is small at low temperature and high density, the quark thermal Green's function approximately renders
\beq
S_W^i\simeq \sum_{\eta=\pm}\frac{\rho^\eta_i}{i\omega_n-(E_p^\eta-\mu_i)},
\eeq
in terms of the density matrices $\rho^\pm_i$ (see Appendix A), where $\omega_n$ denotes the Matsubara frequency, $\omega_n=(2n+1)\pi T$. After some manipulation, we have an expression for the emissivity,
\beqa
\epsilon_{\rm DCDW}&=&2N_cV^{-1}\left[\prod_{i=1}^4 V\int\frac{d^3 p_i}{(2\pi)^3}\right]E_4 W_{fi} n_F(\bp_1)\nonumber\\
&&\times\left(1-n_F(\bp_2)\right)\left(1-n_F(\bp_3)\right),
\label{eq:emis}
\eeqa
where $W_{fi}$ is the transition rate for beta decay of $d$ quark in the DCDW phase. 

\subsection{Transition rate}
\label{subsec:transition}

The transition rate is given as 
\beq
W_{fi}=V(2\pi)^4\delta^{(4)}(p_1-p_2-p_3-p_4\pm q)|M|^2/\prod_{i=1}^4(2E_i V)
\label{eq:wfi}
\eeq
with 
\beq
|M|^2=\frac{1}{2}\sum_{\sigma_1,\sigma_2,\sigma_3}\left|M_{fi}\right|^2,
\label{sqme}
\eeq  
where the squared matrix element can be evaluated as
\begin{eqnarray}
|M_{fi}|^2&=&\frac{{\tilde G}_F^2}{2}{\rm tr}\left(\rho_e \Gamma_\mu\rho_{\nu_e}\Gamma_\nu\right)
{\rm tr}\left(\rho_u \Gamma^\mu\rho_d \Gamma^\nu\right)\nonumber\\
&\equiv& \frac{{\tilde G}_F^2}{2}H^{\mu\nu}L_{\mu\nu}
\end{eqnarray}
in terms of the density matrices, $\rho_i, i=u,d$ for quarks and 
\begin{eqnarray}
\rho_e&=&\ds{p}_3+m_e\nonumber\\
\rho_\nu&=&\ds{p}_4,
\end{eqnarray}
for leptons. Note that the sum over the spin polarizations of quarks is taken in Eq.~(\ref{sqme}).
The leptonic tensor $L_{\mu\nu}$ can be easily evaluated as
\beqa
L_{\mu\nu}&=&\sum_{\sigma_3}{\rm tr}\left(\rho_e \Gamma_\mu\rho_{\nu_e}\Gamma_\nu\right)\nonumber \\
&=& 8\left[p_{3\mu}p_{4\nu}-g_{\mu\nu}p_3p_4+p_{3\nu}p_{4\mu}+i\epsilon_{\alpha\mu\beta\nu} p_3^\alpha p_4^\beta\right] \ . 
\eeqa
The quark tensor $H^{\alpha\beta}$ has a somewhat complicated form.
Consider
\beq
H^{\mu\nu}_{\eta\eta'}\equiv\frac{1}{4}{\rm tr}\left(\Lambda^\eta_u \Gamma^\mu \Lambda^{\eta'}_d \Gamma^\nu \right),
\eeq
for the spin polarization $\eta, \ \eta' (=\pm 1) $ by the use of the density matrices $\Lambda^{\pm}$ in  Appendix A.
The evaluation of the quark tensor is straightforward to give
\beq
H^{\mu\nu}_{\eta\eta'}=2\left[k^{\eta,\mu}_1k^{\eta',\nu}_2-g^{\mu\nu}k^\eta_1k_2^{\eta'}+k^{\eta,\nu}_1k^{\eta',\mu}_2+i\epsilon^{\alpha\mu\beta\nu}k^\eta_{1\alpha}k^{\eta'}_{2\beta}\right]
\eeq
(Appendix B), where $k_i^\eta$ is defined as 
\beqa
k_1^\eta&=&\left(p_1^\eta-\frac{q}{2}\right)\left(1-p_1^\eta Q_1^\eta\right)+m^2Q_1^\eta,\nonumber\\
k_2^{\eta'}&=&\left(p_2^{\eta'}+\frac{q}{2}\right)\left(1-p_2^{\eta'} Q_2^{\eta'}\right)+m^2Q_2^{\eta'},
\eeqa
with $Q_i^\eta=-\eta q/\sqrt{(p_i^\eta q)^2-m^2 q^2}$ and $p_i^\eta=(E_i^\eta,\bp_i)$
, where $\eta$ and $\eta'$ denote the spin polarization for $d$ quark and $u$ quark, respectively.
Then 
\beqa
\left|M^{\eta\eta'}_{fi}\right|^2&=&\frac{{\tilde G}_F^2}{2}H_{\eta\eta'}^{\mu\nu}L_{\mu\nu}\nonumber\\
&=&32{\tilde G}_F^2(k^{\eta'}_2p_3)(k^\eta_1p_4),
\label{matrix2}
\eeqa 
which is reduced to
$
32{\tilde G}_F^2(p_2p_3)(p_1p_4)
$
as $q\rightarrow 0$. After summing over $\sigma_2$ and averaging over $\sigma_1$, we immediately get the Iwamoto's result \cite{iwa80}.

On the other hand, 
\beqa
k_1^\eta&\rightarrow&\left(p_1^\eta-\frac{q}{2}\right)\left(1-\eta\frac{{\bf p}_1\cdot{\bf q}}{|{\bf p}_1\cdot{\bf q}|}\right)\nonumber\\
k_2^{\eta'}&\rightarrow&\left(p_2^{\eta'}+\frac{q}{2}\right)\left(1+\eta' \frac{{\bf p}_2\cdot{\bf q}}{|{\bf p}_2\cdot{\bf q}|}\right),
\label{mless}
\eeqa
in the massless limit. If $\bf q$ is taken as $z$ direction, only the half space is relevant for each momentum integral, depending on $\eta$. 
We shall see the neutrino emission is prohibited in this case, irrespective of $\bf q$, by the energy-momentum conservation unless the interaction is not included, 
as in the direct URCA process. Recalling that the driving mechanism for the emergence of DCDW is the level splitting by the mass term 
between the energy spectra of massless quarks with relative momentum difference $\bf q$. Also, since mass is proportional to the amplitude of DCDW, 
there should not be left any effect in the massless limit.

\section{Phase space integral}
\label{sec:phase}

Taking \cite{max,bah} for references, we try to manipulate the phase space integral for the emissivity (\ref{eq:emis}).
The energy-momentum conservation reads
\beqa
\bp_1&=&\bp_2+\bp_3+\bp_4+\bq,
\label{mom}\\
E_{1}^\eta&=&E_{2}^{\eta'}+E_3+E_4.
\label{ene}
\eeqa
Dropping $\bp_4$ in Eq.~(\ref{mom}) because of $|\bp_4|=O(T)$,
\beq
\left(\bp_1-\bq/2\right)^2\simeq \left(\bp_2+\bq/2\right)^2+|\bp_3|^2+2\bp_3\left(\bp_2+\bq/2\right),
\label{mom2}
\eeq
which is recast into
\beqa
&&(E_{1}^{\eta})^2-\left(\eta\sqrt{(\bp_1\cdot \bq)^2+m^2|\bq|^2}+\bq\cdot\bp_1\right)\nonumber\\
&\simeq& (E_{2}^{\eta'})^2-\left(\eta' \sqrt{(\bp_2\cdot \bq)^2+m^2|\bq|^2}-\bq\cdot\bp_2\right)+\nonumber\\
&+&|\bp_3|^2+2\bp_3\left(\bp_2+\bq/2\right) \ , 
\label{mom3}
\eeqa
where we put $m_u\simeq m_d\equiv m$. Similarly, we find 
\beq
(E_{1}^{\eta})^2\simeq (E_{2}^{\eta'})^2+{E_3}^2+2E_{2}^{\eta'}E_3,
\label{ene2}
\eeq
from Eq.~(\ref{ene}) by neglecting $E_4$ again. From Eqs.~(\ref{mom3}), (\ref{ene2}) we have
\beqa
&&\left(\eta\sqrt{(\bp_1\cdot \bq)^2+m^2|\bq|^2}+\bq\cdot\bp_1\right)\nonumber\\
&-&\left(\eta' \sqrt{(\bp_2\cdot \bq)^2+m^2|\bq|^2}-\bq\cdot\bp_2\right)\nonumber\\
&+&2\bp_3\left(\bp_2+\bq/2\right)
\simeq 2E_3E_2^{\eta'},
\label{cons}
\eeqa
where we used $E_3\simeq |\bp_3|$.

\subsection{Case of the massless-quark limit}
\label{subsec:massless}

First, we consider the massless limit by setting $m=0$. Eq.~(\ref{cons}) is then reduced to a simple one,
\beqa
&&\eta |\bp_1\cdot\bq|\left(1+\eta\frac{\bp_1\cdot\bq}{|\bp_1\cdot\bq|}\right)-\eta' |\bp_2\cdot\bq|\left(1-\eta' \frac{\bp_2\cdot\bq}{|\bp_2\cdot\bq|}\right)\nonumber\\
&+&2\bp_3\left(\bp_2+\bq/2\right)\simeq 2E_3E_2^{\eta'}.
\label{nogo}
\eeqa
The first two terms should be vanished for the non-zero value of $k_i^\alpha$ from Eq.~(\ref{mless}).
Using Eq.~(\ref{mless}), the squared matrix element (\ref{matrix2}) is also reduced to a simple one,
\beqa
\left|M^{\eta\eta'}_{fi}\right|^2&=&32{\tilde G}_F^2(k_1^\eta p_4)\left(1+\eta' \frac{\bp_2\cdot\bq}{|\bp_2\cdot\bq|}\right)\nonumber\\
&&\times\left(E_{p_2}^{\eta'} E_3-\left(\bp_2+\bq/2\right)\cdot \bp_3\right),
\label{msq}
\eeqa
which gives no contribution due to Eqs.~(\ref{mless}) and (\ref{nogo}).
Generally the emissivity is vanished as $\bq$ or $m$ goes to zero as it should be.

Since the Fermi surface is well deformed as the wave vector $\bq$ increases \cite{dcdw}, the general expression of the emissivity is difficult to be evaluated. However, one may estimate it by considering the specific cases near the phase boundaries of the DCDW phase, where the deformation is very weak at one side and extremely strong at the other side. 

\subsection{Near the onset density of DCDW}
\label{subsec:onset}

\subsubsection{Effective Fermi sphere}
\label{subsubsec:fermi}

First, we consider the cooling rate near the onset density of the DCDW phase, where the deformation of the Fermi surface is not so remarkable (Fig.~1). So, one may introduce the effective Fermi sphere instead of the realistic Fermi surface, keeping the volume fixed. The Fermi sphere of the minor spins is already sufficiently small, and we can safely discard its contribution (one Fermi sea approximation); we, hereafter, only consider the $u,d$ quarks with $\eta=\eta'=-1$. Moreover, since the dynamical quark mass is rather small compared with the quark chemical potentials, $m_u\simeq m_d\ll\mu_i$, we may treat them as massless quarks. 
The volume of each Fermi sphere can be easily evaluated,
\beqa
V_F^i&=&2\pi\int^{p_z^{\rm max}}_0 dp_z\left[|\bq|\sqrt{m^2+p_z^2}-p_z^2+\mu_i^2-m^2-|\bq|^2/4\right]\nonumber\\
&=&2\pi\left[|\bq|/2\left(p_z^{\rm max}\sqrt{p_z^{\rm max,2}+m^2}\right.\right.\nonumber\\
&+&\left.m^2{\rm ln}\left(\frac{p_z^{\rm max}+\sqrt{p_z^{\rm max,2}+m^2}}{m}\right)\right)\nonumber\\
&&\left.-\frac{p_z^{\rm max, 3}}{3}+\left(\mu_i^2-m^2-|\bq|^2/4\right)p_z^{\rm max}\right],
\eeqa
where $p_z^{\rm max}\simeq \sqrt{(\mu_i+|\bq|/2)^2-m^2}$ for each $u$ or $d$ quark. Thus the radius of the effective Fermi sphere is given as
\beq
{\bar p}_{Fi}\simeq\left(\mu_i+|\bq|/2\right)\left[\frac{4\mu_i-|\bq|}{4\mu_i+2|\bq|}\right]^{1/3}\simeq \mu_i+|\bq|/4+...,
\eeq
where we have used $\mu_i\gg |\bq|/2\simeq m$. Note that the quark energy is now approximated as $E^\pm_\bp\simeq |\bp|$ within the same approximation. In the following we evaluate the emissivity by assuming massless quarks in the presence of DCDW. Usually it vanishes in the absence of DCDW by the kinematical conditions. Following Iwamoto \cite{iwa80}, we begin with the formula,
\beq
|M_{fi}^{\eta\eta'}|^2=32{\tilde G}_F^2(p_1\cdot p_4)(p_2\cdot p_3),
\eeq
for the squared matrix element. Note that the factor 2 is different from \cite{iwa80} since the only one polarization is relevant. 
It can be further written as
\beq
|M_{fi}^{\eta\eta'}|^2\simeq 32{\tilde G}_F^2E_1E_2E_3E_4(1-\cos\theta_{14})(1-\cos\theta_{23}),
\label{eq:matrixonset}
\eeq
where $\theta_{14}$ ($\theta_{23}$) is the angle between momenta of the $d$ quark and neutrino (the angle between momenta of the $u$ quark and electron).  

\subsubsection{Expression of the emissivity}
\label{subsubsec:emisonset}

Setting the momentum magnitudes of quarks and electrons equal to their values on the respective Fermi surfaces, one obtains
\beqa
&&\epsilon_{\rm DCDW}\simeq\frac{3}{(2\pi)^8}{\bar p}^2_{Fu}{\bar p}^2_{Fd}\mu_e^2\int dE_1dE_2dE_3E_4^2dE_4\nonumber\\
&\times&\delta(E_1-E_2-E_3-E_4)\prod_{i=1}^3(2E_i)^{-1} \nonumber\\
&\times& n(\bp_1)[1-n(\bp_2)][1-n(\bp_3)] \left(\prod_{i=1}^4 \int d\Omega_i\right) |M_{fi}^{\eta\eta'}|^2 \nonumber\\
&\times& \delta^{(3)}(\bp_1-\bp_2-\bp_3-\bp_4-\bq),
\label{eq:emisonset}
\eeqa
where only one polarization has been taken into account for the initial $d$ quarks.
First we can proceed the angular integral as in the pion-condensed case\cite{max}: 
\begin{equation}
A\equiv \prod_{i=1}^4\left(\int d\Omega_i \right) |M_{fi}^{\eta\eta'}|^2 \delta^{(3)}(\bp_1-\bp_2-\bp_3-\bp_4-\bq) \ .
\label{eq:angleonset}
\end{equation}
In the following, the neutrino momentum $\bp_4$ in the delta function is dropped because of $|\bp_{\bar\nu}|=O(T)$. After integrating with respect to the angle $\Omega_4$ with the squared matrix element (\ref{eq:matrixonset}), one has
\begin{equation}
 A\simeq\frac{32}{2\pi^2}\tilde G_F^2E_1E_2E_3E_4\tilde A \ , 
 \label{eq:angleonset2}
 \end{equation}
 where
\begin{eqnarray}
\tilde A&=&\left(\prod_{i=1}^3 \int d\Omega_i \right)(1-\cos\theta_{23})\int d^3 x e^{i(\bp_1-\bp_2-\bp_3- \bq)\cdot x} \cr
 &\simeq&\frac{64\pi^5}{|{\bp_1}||{\bp_2}||\bq|}.
 \label{eq:angleonset3}
\end{eqnarray}
The derivation of the $\tilde A$ is given in Appendix~\ref{sec:c}. 
Note that this integral gives a finite value only if the triangle condition is satisfied in the limit $|\bq|\rightarrow 0$ as shown in Appendix C. The remaining phase-space integration in Eq.~(\ref{eq:emisonset}) leads to the emissivity, 
\beq
\epsilon_{\rm DCDW}\simeq \frac{3}{(2\pi)^5}\mu_u\mu_d\frac{\mu_e^2}{2|\bq|}32{\tilde G}_F^2I,
\eeq
with
\beqa
I&=&\left(\prod_{i=1}^3\int_{-\infty}^{\infty} dE_i\right)\int_0^\infty dE_4E_4^3\delta(E_1-E_2-E_3-E_4)\nonumber\\
&&\times n(\bp_1)[1-n(\bp_2)][1-n(\bp_3)]=\frac{457}{5040}\pi^6T^6.
\label{int}
\eeqa
The emissivity of the neutrino process, $u+e^-\rightarrow d+\nu_e$, gives the same contribution as that of the process, $d\rightarrow u+e^-+\bar\nu_e$. Therefore, by multiplying a factor 2, one finally has
\beq
\epsilon_{\rm DCDW}\simeq \frac{457}{1680}\pi {\tilde G}_F^2 \mu_u\mu_d\frac{\mu_e^2}{|\bq|}T^6.
\label{eq:emisonsetfinal}
\eeq
Assuming $\mu_u=\mu_d$, as in the non-interacting $u,d$ quark matter, and  using the values in Table \ref{para}, we can estimate its numerical value as
\beq
\epsilon_{\rm DCDW}\simeq 6.1\times 10^{26}(\rho_B/\rho_0)^{2/3}Y_e^{2/3}T_9^6~~~({\rm erg\cdot cm^{-3}\cdot s^{-1}}),
\label{eq:emisonsetnum}
\eeq
where $Y_e$ is the electron number fraction in quark matter, $Y_e=\rho_e/\rho_B$, $\rho_0$ the nuclear saturation density, $\rho_0\simeq 0.17$fm$^{-3}$, and $T_9\equiv T/10^9({\rm K})$.

\subsection{Near the termination density of DCDW}
\label{subsec:weak}

\subsubsection{Deformation of the Fermi surface}
\label{subsubsec:deform}

\begin{figure}[htbp]
 \begin{center}
  \subfigure{\includegraphics[height=0.45\linewidth]{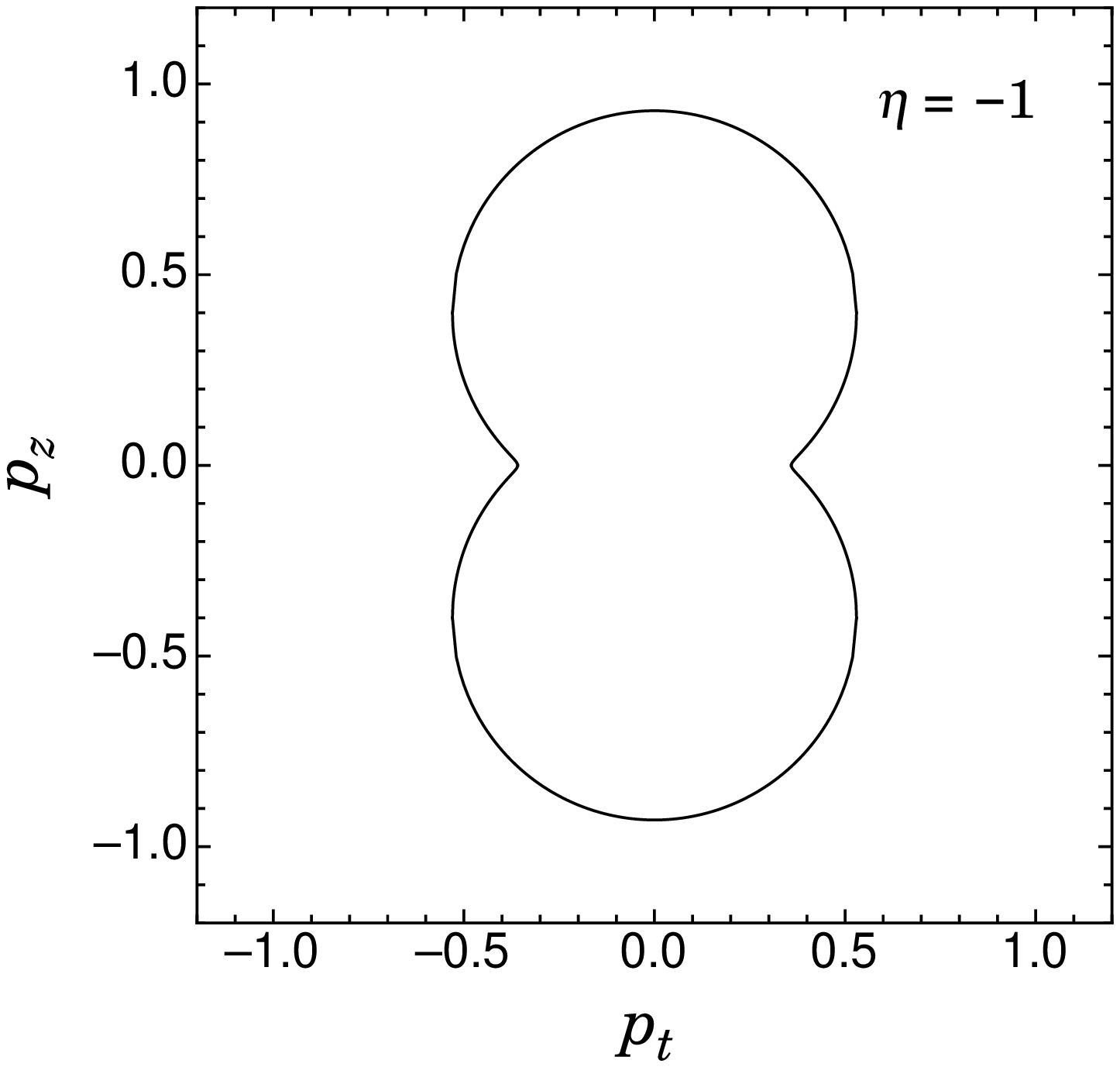}}
  \hspace{15mm}
  \subfigure{\includegraphics[height=0.45\linewidth]{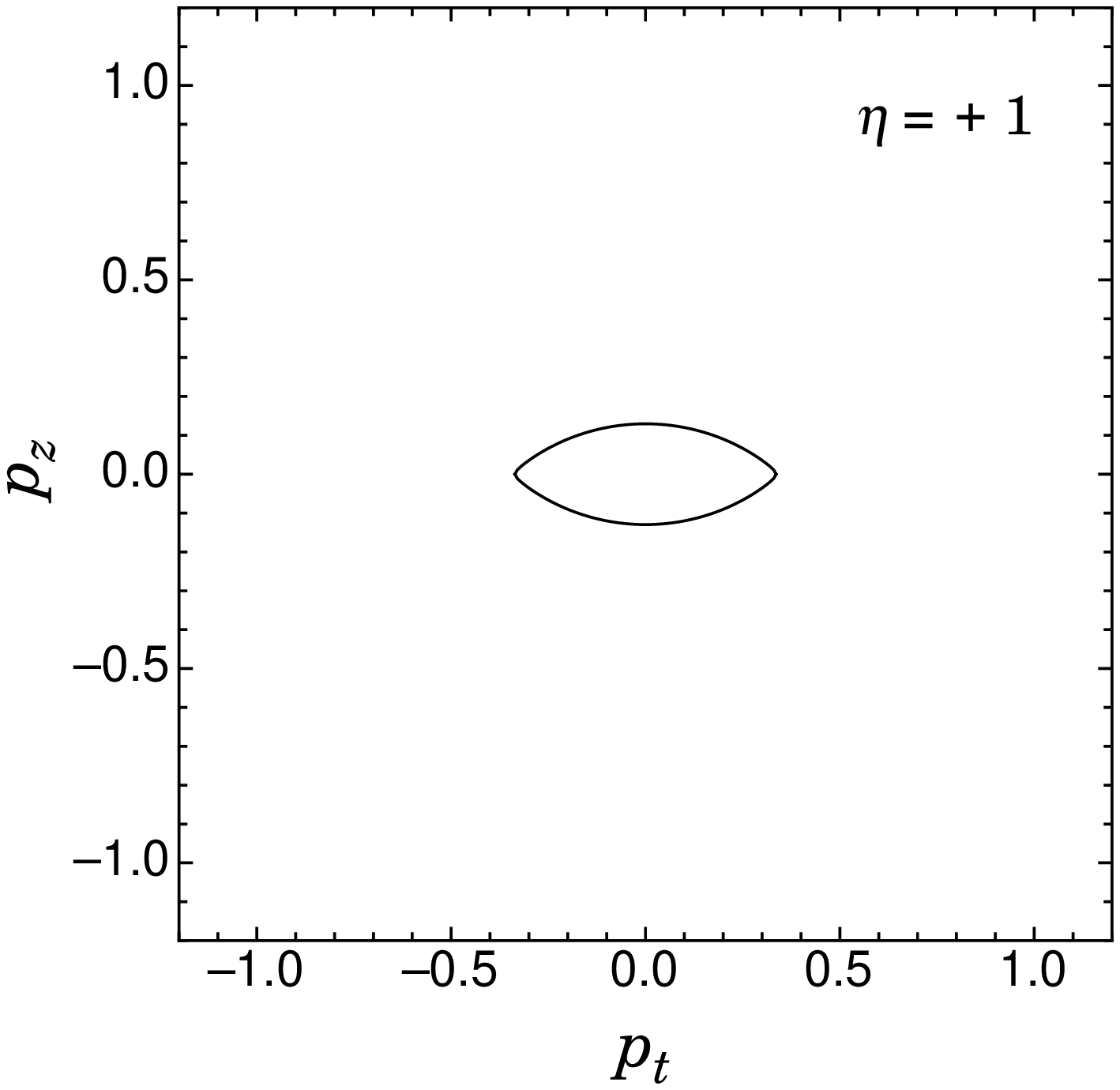}}
 \caption{Fermi surfaces at the termination density with arbitrary scale.  The meaning of the legend and symbols are the same as in Fig.~\ref{fig1}.}
 \label{fig2}
\end{center}
\end{figure}

Near the termination density the dynamical quark mass is also very small while the wave vector $\bf q$ is still large. 
The quark energy is well approximated as $E^-_p\simeq |\bp\pm\bq/2|$ for major quarks:
the Fermi seas are then remarkably deformed to be almost separated spheres with centers shifted by $\bq$ (see Fig.~2). 
Using Eq.~(\ref{cons}), we have
the squared matrix element (\ref{matrix2}),
\begin{eqnarray}
 &&|M_{fi}^{\rm \eta\eta'}|^2\simeq 32\tilde G_F^2\Bigg\lbrack 
\Big\lbrace E_1E_4-\left({\bf p}_1-{\bf q}/2\right)\cdot {\bf p}_4\Big\rbrace \cr
&\times&\Bigg\lbrace\frac{\eta}{2}\frac{m^2|{\bf q}|^2}{\sqrt{( {\bf p}_1\cdot{\bf q})^2+m^2|{\bf q}|^2}}\left(1+\eta' \frac{{\bf p}_2\cdot{\bf q}}{\sqrt{( {\bf p}_2\cdot{\bf q})^2+m^2|{\bf q}|^2}}\right) \cr
& -&\eta' \frac{m^2({\bf p}_3+{\bf q}/2)\cdot{\bf q}}{\sqrt{( {\bf p}_2\cdot{\bf q})^2+m^2|{\bf q}|^2}} \left(1-\eta\frac{{\bf p}_1\cdot{\bf q}}{\sqrt{( {\bf p}_1\cdot{\bf q})^2+m^2|{\bf q}|^2}}\right)\Bigg\rbrace \cr
&+&\frac{1}{2}m^2({\bf p}_4\cdot{\bf q})
 \left(1+\eta\frac{{\bf p}_1\cdot{\bf q}}{\sqrt{( {\bf p}_1\cdot{\bf q})^2+m^2|{\bf q}|^2}}\right) \cr
 &\times&\left(1+\eta' \frac{{\bf p}_2\cdot{\bf q}}{\sqrt{( {\bf p}_2\cdot{\bf q})^2+m^2|{\bf q}|^2}}\right) \cr
 &-&\eta\eta' \frac{m^4 ({\bf p}_3+{\bf q}/2)\cdot{\bf q} \ ({\bf p}_4\cdot {\bf q})}{\sqrt{( {\bf p}_1\cdot{\bf q})^2+m^2|{\bf q}|^2} \sqrt{( {\bf p}_2\cdot{\bf q})^2+m^2|{\bf q}|^2}} 
 \Bigg\rbrack 
\label{eq:mat4}
\end{eqnarray}
with the proper momentum restriction. Note that it is obviously vanished when $m\rightarrow 0$ or $|\bq|\rightarrow 0$.

Next consider the angular integrations of the squared matrix element. Since one spin polarization is dominant, it is sufficient to consider only the case where $\eta=\eta'=-1$ for $u,d$ quarks.

\begin{figure}[h]
\begin{center}
\includegraphics[height=.20\textwidth]{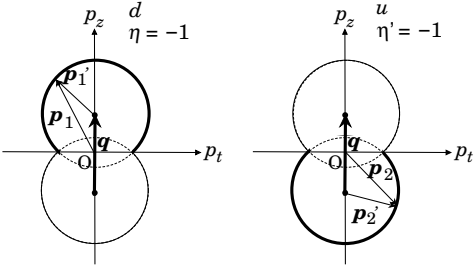}
\caption{Schematic view of the Fermi surfaces for $d$ quarks with $\eta=-1$ and $u$ quarks with $\eta' = -1$ near the termination density. See the text for the meaning of the legend and symbols.} 
\label{fig3}
\end{center}
\end{figure}

We first perform the angular integration in (15),
\beq
B=\left(\prod_{i=1}^4\int d\Omega_i\right)\left|M^{\eta\eta'}_{fi}\right|^2\delta^{3}\left(\bp_1-\bp_2-\bp_3-\bp_4-\bq\right).
\label{eq:B}
\eeq
Near the termination density, $m\ll|{\bf p}_i| \simeq\mu_i <|{\bf q}|$.  ($i=u,d$) [see Table~\ref{para}]. Thereby we make an approximation to neglect the terms $m^2|{\bf q}|^2$ appearing in the denominators in the four parentheses ($\cdots$) in Eq.~(\ref{eq:mat4}). 
Furthermore we drop $\bp_4$ from the delta function in Eq.~(\ref{eq:B}) since $|\bp_4|=O(T)$.
Then, substituting the matrix element (\ref{eq:mat4}) into (\ref{eq:B}) and after integrating with respect to the angle $\Omega_4$, one obtains
\begin{eqnarray}
&B&=\frac{32}{2\pi^2}\tilde G_F^2 m^2 E_1^\eta E_4 \left(\prod_{i=1}^3\int d\Omega_i\right)\int d^3 x
e^{i({\bf p}_1-{\bf p}_2-{\bf p}_3-{\bf q})\cdot{\bf x}} \cr
&\times&\Bigg\lbrack \frac{|{\bf q}|^2}{2}\Bigg\lbrace -\frac{1}{\sqrt{({\bf p}_1\cdot {\bf q})^2+m^2|{\bf q}|^2}}\left(1-\frac{{\bf p}_2\cdot{\bf q}}{|{\bf p}_2\cdot{\bf q}|}\right) \cr
&+&\frac{1}{\sqrt{({\bf p}_2\cdot {\bf q})^2+m^2|{\bf q}|^2}}\left(1+ \frac{{\bf p}_1\cdot{\bf q}}{|{\bf p}_1\cdot{\bf q}|}\right) \Bigg\rbrace\cr
&+&\frac{{\bf p}_3\cdot{\bf q}}{\sqrt{({\bf p}_2\cdot{\bf q})^2+m^2|{\bf q}|^2}}\left(1+\frac{{\bf p}_1\cdot{\bf q}}{|{\bf p}_1\cdot{\bf q}|}\right)\Bigg\rbrack \ .
\label{eq:B1}
\end{eqnarray}
The available range of the momentum ${\bf p}_1$ and ${\bf p}_2$ contributing to the $B$ is such that ${\bf p}_1\cdot {\bf q} > 0$ and ${\bf p}_2\cdot {\bf q} < 0$, which result from $\left(1+{\bf p}_1\cdot{\bf q}/|{\bf p}_1\cdot{\bf q}|\right)>0$ and $\left(1-{\bf p}_2\cdot{\bf q}/|{\bf p}_2\cdot{\bf q}|\right) > 0$ in Eq.~(\ref{eq:B1}). As illustrated in Fig.~\ref{fig3}, only the $d$ quarks occupying the upper part of the ``two-center'' Fermi surface and $u$ quarks occupying the lower part of the ``two-center'' Fermi surface contribute to the reaction. 
Changing the variables $\bp_1,\bp_2$ by the new ones,
\beq
\bp'_1\equiv \bp_1-\bq/2, ~~~\bp'_2\equiv \bp_2+\bq/2,
\eeq
we have, from Eq.~(\ref{eq:energy}), 
\beq
E_1^\eta\simeq |\bp_1'|,~~~E_2^{\eta'}\simeq |\bp_2'|,
\eeq
which means both the angular integrations with respect to $\bp'_i$ ($i$=1, 2) have spherical symmetry. In terms of the new variables ${\bf p}_1'$, ${\bf p}_2'$, $B$ is rewritten as
\begin{equation}
B=\frac{32}{2\pi^2}\tilde G_F^2 m^2 E_1^\eta E_4(\tilde B_1+\tilde B_2+\tilde B_3) \ , 
\label{eq:B123}
\end{equation}
where
 \begin{subequations}\label{eq:Btilde}
\begin{eqnarray}
\tilde B_1&=&-\left(\prod_{i=1}^3\int d\Omega_i\right)\int d^3 x \exp\left[i({\bf p}_1'-{\bf p}_2'-{\bf p}_3)\cdot{\bf x}\right]\cr
&\times&\frac{|{\bf q}|^2}{\sqrt{\left\{({\bf p}_1'+{\bf q}/2)\cdot {\bf q}\right\}^2+m^2|{\bf q}|^2}} \cr
&\simeq&-32\pi^5\frac{|{\bf q}|}{|{\bf p}_1'|^2 |{\bf p}_2'||{\bf p}_3|}\log\left(\frac{2|{\bf q}|}{m}\right) \ , \label{eq:Btilde1} \\
\tilde B_2&=&\left(\prod_{i=1}^3\int d\Omega_i\right)\int d^3 x \exp\left[i({\bf p}_1'-{\bf p}_2'-{\bf p}_3)\cdot{\bf x}\right] \cr
&\times&\frac{|{\bf q}|^2}{\sqrt{\left\{({\bf p}_2'-{\bf q}/2)\cdot {\bf q}\right\}^2+m^2|{\bf q}|^2}} \cr
&\simeq& 32\pi^5\frac{|{\bf q}|}{|{\bf p}_1'| |{\bf p}_2'|^2|{\bf p}_3|}\log\left(\frac{2|{\bf q}|}{m}\right) \ , \label{eq:Btilde2}  \\
\tilde B_3&=&\left(\prod_{i=1}^3\int d\Omega_i\right)\int d^3 x \exp\left[i({\bf p}_1'-{\bf p}_2'-{\bf p}_3)\cdot{\bf x}\right] \cr
&\times&\frac{2({\bf p}_3\cdot{\bf q})}{\sqrt{\left\{({\bf p}_2'-{\bf q}/2)\cdot {\bf q}\right\}^2+m^2|{\bf q}|^2}} \cr
&\simeq& 16\pi^5|{\bf q}|\frac{|{\bf p}_1'|^2-|{\bf p}_2'|^2-|{\bf p}_3|^2}{|{\bf p}_1'| |{\bf p}_2'|^4|{\bf p}_3|} \cr
&\times&\left[\log\left(\frac{2|{\bf q}|}{m}\right)-2\right]
\ . \label{eq:Btilde3}
\end{eqnarray}
\end{subequations}
 The details of evaluating the $\tilde B_i$ ($i=1-3$) are shown in Appendix~\ref{sec:d}.
 
\subsubsection{Expression of the emissivity}
\label{subsubsec:emisterminal}

By the use of Eqs.~(\ref{eq:B123}) and (\ref{eq:Btilde}) for the angular integral $B$, the emissivity [Eq.~(\ref{eq:emis})] is written in the case of the region near the termination density as 
\beqa
\epsilon_{\rm DCDW}&\simeq&\frac{3}{(2\pi)^8}|{\bf p}_{Fu}|^2 |{\bf p}_{Fd}|^2\mu_e^2\int dE_1dE_2dE_3E_4^2dE_4\nonumber\\
&\times&\delta(E_1-E_2-E_3-E_4)\prod_{i=1}^3(2E_i)^{-1} \nonumber\\
&\times& n(\bp_1)[1-n(\bp_2)][1-n(\bp_3)] B \ ,
\label{eq:emisterminal}
\eeqa
where $|{\bf p}_{Fi}|$ ($i=u,d$) is the Fermi momentum of the quark. 
Noting that $|{\bf p}_1'|\sim E_1$, $|{\bf p}_2'|\sim E_2$, $|{\bf p}_3|\sim E_3$, $|{\bf p}_4|=E_4$ and by the use of  Eq.~(\ref{int}), one can perform the remaining phase-space integrations in (\ref{eq:emisterminal}). With the help of the chemical equilibrium relation,
\begin{equation}
\mu_d=\mu_u+\mu_e \ , 
\label{eq:chem}
\end{equation}
and by multiplying a factor 2 to take into account the neutrino process, $u+e^-\rightarrow d+\nu_e$, one obtains the final expression for the emissivity:
\begin{eqnarray}
&&\epsilon_{\rm DCDW}=\frac{1}{2}\frac{457}{1680}\pi \tilde G_F^2 m^2|{\bf q}|\frac{\mu_e}{\mu_u} T^6\cr
&\times&\left\{\log\left(\frac{2|{\bf q}|}{m}\right)+\frac{\mu_d}{\mu_u}\left[\log\left(\frac{2|{\bf q}|}{m}\right)-2\right] \right\} \ . 
\label{eq:emisfinal}
\end{eqnarray}
Assuming again $\mu_u\simeq\mu_d$ in the $u,d$ quark matter, and using the values in Table \ref{para}, we can estimate the numerical value,
\begin{equation}
\epsilon_{\rm DCDW}=(2.16\times 10^{24})\left(\frac{\rho_{\rm B}}{\rho_0}\right)^{1/3}Y_e^{1/3}T_9^6 \quad ({\rm erg}\cdot {\rm cm}^{-3}\cdot {\rm s}^{-1} ) \ . 
\label{eq:emisn2}
\end{equation}

\section{Discussion}
\label{sec:discussion}

In both regions near the onset density (I) and near the termination density (II), the wave vector ${\bf q}$, which marks inhomogeneity of the DCDW phase, plays an essential role on enhancement of neutrino emissions via the quark beta-decay. Owing to the existence of ${\bf q}$, there is no need to supply energy and momentum to the reactions through spectator particles. As a result, the available phase space for the quark beta decay in the DCDW phase is enough to give a large neutrino emissivity which is proportional to $T^6$, as is the case with other exotic cooling mechanisms\cite{iwa80,max,mut,tat}. 

The neutrino emissivity in the DCDW phase near the onset density, $\epsilon_{\rm DCDW}^{(\rm I)}$, is proportional to $1/|{\bf q}|$ [ (\ref{eq:emisonsetfinal}) ]. This ${\bf q}$-dependence originates from angular integral of the phase factor, $\exp(- i{\bf q}\cdot {\bf x})$ in (\ref{eq:angleonset3}), and such specific momentum-dependence is similar to that in the pion-condensed case, where the neutrino emissivity, given on the basis of spherical Fermi surfaces for baryons, is proportional to $1/|{\bf k}|$ with $|{\bf k}|$ being the momentum of $p$-wave pion condensates\cite{max,mut}. The magnitude of $\epsilon_{\rm DCDW}^{(\rm I)}$ [ (\ref{eq:emisonsetnum}) ] is of the same order as those for normal quark cooling\cite{iwa80} and pion cooling\cite{max,mut}. On the other hand, the neutrino emissivity in the DCDW phase near the termination density, $\epsilon_{\rm DCDW}^{(\rm II)}$, has a complicated dependence on $|{\bf q}|$ and $m$ including the terms $\propto m^2|{\bf q}|\log\left(2|{\bf q}|/m\right)$ [ (\ref{eq:emisfinal}) ],  
which reflects a singular structure of the quark tensor originating from the deformed Fermi surface [see Figs.~\ref{fig2} and \ref{fig3}]. The resulting emissivity $\epsilon_{\rm DCDW}^{(\rm II)}$ is smaller than $\epsilon_{\rm DCDW}^{(\rm I)}$ by two orders of magnitude [see (\ref{eq:emisonsetnum}) and (\ref{eq:emisn2})], but still larger than emissivities for standard cooling processes such as the modified Urca process by a factor $\sim 10^3/ T_9^2$~\cite{fm79}.

 It should be noted that the enhancement of neutrino emissions works in the limited density region, because the DCDW phase appears only 
near the chiral transition. Only the shell region of the radius width $\Delta R$ inside hybrid stars is responsible to the fast cooling mechanism. If enhanced cooling region is limited only to such inhomogeneous phases, heavier compact stars may not necessarily cool faster than lighter ones. This opens up another possibility for explaining the thermal evolution of Cas A and other cooler stars in a consistent way, as recently proposed by Noda et al. based on the model separating quark matter region into the CFL phase and non-superconducting quark phase\cite{nod}. 

\section{Summary and concluding remarks}
\label{sec:summary}

We have proposed a novel cooling mechanism (DCDW cooling) of hybrid stars, based on the idea of the inhomogeneous chiral phase.
It originates from the non-perturbative effect of QCD at moderate densities.
We have shown that the beta decay process becomes possible in the DCDW phase due to the momentum supply by DCDW 
at the weak-interaction vertex. 
The emissivity is estimated near the phase boundaries of the DCDW phase to be the order of $10^{24-26}T_9^6$ (erg$\cdot$ cm$^{-3}$s$^{-1}$), 
which value may be comparable with that by the quark cooling \cite{iwa80} or pion cooling \cite{max,mut}. 
Another important point is that this mechanism works in the limited density region where the DCDW phase appears.
This feature is similar to the Cooper pair-breaking-formation (PBF) process, 
where the limited density region is efficient in the superfluid phase \cite{pbf}.

If we incorporate this mechanism in the calculation of the cooling curves of young neutron stars, 
further works are needed which consider the realistic equation of state (EOS) of cold catalyzed quark matter instead of 
flavor symmetric quark matter and the numerical values of emissivity over the whole region of the DCDW phase without 
the restriction to the phase boundaries \cite{mar}. The effects of the magnetic field is also an interesting subject, 
since there should be large magnetic field inside compact stars.
The appearance of DCDW looks to be robust and less model-dependent 
in the presence of the magnetic field \cite{fro,tat2,nis}.

In this paper we considered the DCDW phase as a typical inhomogeneous chiral phase, 
but the similar mechanism may be possible for other configurations such as RKC. 

It is also interesting to seek for other phenomenological implications of the inhomogeneous chiral phases 
by considering their elasticity \cite{sot} or magnetic properties.

\section*{Acknowledgement}

The authors thank T. Maruyama, N. Yasutake and T. Noda for their interest in this work. This work is partially supported 
by the Grant-in-Aid for Scientific Research on Innovative Areas 
through No. 24105008 provided by MEXT and the Grant-in-Aid for Scientific Research (C) (No. 23540318) by JSPS.

\appendix
\section{Quark propagator in the DCDW phase}
\label{sec:a}

The quark propagator is given by 
\beq
S_W^i(p)=\frac{1}{\ds{p}-m+\gamma_5\tau_3\ds{q}/2}\equiv \frac{N^i}{D},
\eeq
with $i=u,d$ for $\tau_3=\pm 1$, respectively, in the DCDW phase within the mean-field approximation \cite{dcdw}, 
where the index $W$ indicates that we define the new quark field by way of the Weinberg transformation from the original one,
\beq
\psi_W\equiv {\rm exp}\left(i\gamma_5\tau_3\bq\cdot\br/2\right)\psi.
\eeq
The numerator $N^i$ is 
\beq
N^i=(\ds{p}+m-\gamma_5\tau_3\ds{q}/2)(p^{2}-m^{2}+q^2/4
-(pq-m\ds{q})\gamma_5\tau_3),
\eeq
and the denominator $D$ is 
\beq
D=(p^{2}-m^{2}+q^2/4)^2-((pq)^2-m^{2}q^2).
\eeq
The solutions for $D=0$, which is a transcendental equation, 
 give the four energies corresponding to
positive and negative solutions with two polarizations $\eta=\pm
1$: the positive energy solutions are given by 
\beq
E_p^{\pm}=\sqrt{E_{p}^{2}+|{\bf q}|^2/4\pm \sqrt{({\bf
p}\cdot{\bf q})^2+m^{2}|{\bf q}|^2}},
\eeq
with $E_{p}=(m^{2}+|{\bf p}|^{2})^{1/2}$.

The density matrix for the positive-energy state is then given as
\beqa
\rho_i^\pm&=&{\rm Res} S_W^i(p)|_{p_0=E^{\pm}_p}\cr
&=&\frac{(\ds{p}+m-\gamma_5\tau_3\ds{q}/2)|_{p_0=E^\pm_p}(1\pm{\hat s}(p)\tau_3)}{4E^\pm_p}\nonumber\\
&\equiv&\frac{\Lambda^\pm_i}{4E^\pm_p},
\label{eq:dm}
\eeqa
with ${\hat s}(p)\equiv(\bp\cdot\bq+m\ds{q})\gamma_5/\sqrt{(\bp\cdot\bq)^2+m^2|\bq|^2}, {\hat s}^2(p)=1$.
We can easily check 
\beq
\sum_{p_0=E^{\pm}_p}{\rm Res} S_W^i(p)
\rightarrow \frac{\ds{p}+m}{2E_p},
\eeq
as should be in the limit, $\bq\rightarrow 0$. Thus the $i$ quark propagator can be written as 
\beq
S_W^i\simeq \sum_{\eta=\pm}\frac{\rho^\eta_i}{i\omega_n-(E_p^\eta-\mu_i)},
\eeq
once only the positive-energy state is relevant.

\section{Quark tensor}
\label{sec:b}

We calculate the quark tensor $H_{\eta\eta'}^{\mu\nu}$ which is given by
\begin{equation}
H_{\eta\eta'}^{\mu\nu}
=\frac{1}{4}{\rm tr}\Big\lbrack\Lambda_u^{\eta'}\gamma^\mu(1-\gamma_5)\Lambda_d^{\eta}\gamma^\nu(1-\gamma_5)\Big\rbrack \ . 
\label{eq:qtensor1}
\end{equation}
From Eq.~(\ref{eq:dm}) in Appendix A, the density matrix $\Lambda_i^\pm$ for quark ($i=u,d$) is written as
\begin{equation}
\Lambda_i^\eta = \left( {\ds p}_i^\eta+m-\gamma_5\frac{\tau_3}{2}{\ds q} \right)\left[1+\eta\frac{(\bp_i\cdot \bq+m\ds q)\gamma_5\tau_3}{\sqrt{(\bp_i\cdot \bq)^2+m^2|\bq|^2}}\right] \ ,
\label{eq:lamu}
\end{equation}
where $\eta=\pm 1$, and $\tau_3$=1 ($\tau_3=-1$) for $i=u$ ($i=d$). 
The four-vectors, $p_i^\eta$ and $q$, are represented as $p_i^\eta=(E_{p_i}^{\eta}, \  \bp_i)$ and 
$q^\alpha=(0, \bq)$, respectively. 
After manipulation with the Dirac matrices, one obtains
\beqa
\Lambda_i^\eta &=&\left(m+\Ds A_i^{\eta} -\frac{1}{2}\frac{\eta m}{\sqrt{(\bp_i\cdot \bq)^2
+m^2|\bq|^2}}\ds q \ds q \right)\nonumber\\
&&+\left(B_i^{\eta}+\Dds C_i^{\ \eta}+\eta\frac{m\tau_3 {\ds p}_i^\eta\ds q}{\sqrt{(\bp_i\cdot \bq)^2+m^2|\bq|^2}}\right)\gamma_5 \ , \qquad  
\label{eq:lam}
\eeqa
where 
${\Ds  A}_i^\eta=\gamma^\mu (A_i^\eta)_\mu$, ${\Dds C}_i^{\ \eta}=\gamma^\mu (C_i^\eta)_\mu$, and
\begin{subequations}\label{eq:abc2}
\begin{eqnarray}
A_i^{\eta}&\equiv& p_i^\eta+\eta \frac{1}{2} q\frac{\bp_i\cdot \bq }{\sqrt{(\bp_i\cdot \bq)^2+m_i^2|\bq|^2}} \ ,
\label{eq:abca2} \\
B_i^{\eta}&\equiv&\eta m\frac{\bp_i\cdot \bq \ \tau_3}{\sqrt{(\bp_i\cdot \bq)^2+m^2|\bq|^2}} \ , 
\label{eq:abcb2} \\
C_i^{\eta}&\equiv&\eta p_i^\eta\frac{\bp_i\cdot \bq \ \tau_3}{\sqrt{(\bp_i\cdot \bq)^2+m^2|\bq|^2}} +\eta\frac{m^2 q \ \tau_3}{\sqrt{(\bp_i\cdot \bq)^2+m^2|\bq|^2}} \cr
&+&\frac{\tau_3}{2}q \ ,
\label{eq:abcc2}
\end{eqnarray}
\end{subequations}

Substitution of Eq.~(\ref{eq:lam}) into $\Lambda_u^{\eta'}$ and $\Lambda_d^{\eta}$ on the r.h.s. of Eq.~(\ref{eq:qtensor1}) leads to
 \begin{equation}
H_{\eta\eta'}^{\mu\nu}=\frac{1}{2}{\rm tr}\Big\lbrack(\Ds A_u^{\eta'}+\Dds C_u^{\ \eta'})
\gamma^\mu (\Ds A_d^{\eta}+\Dds C_d^{\ \eta})\gamma^\nu(1-\gamma_5)\Big\rbrack \ . 
\label{eq:qtensor4}
\end{equation}
 By the use of the formulae, ${\rm tr}(\gamma^\mu\gamma^\nu\gamma^\rho\gamma^\sigma)=4(g^{\mu\nu}g^{\rho\sigma}-g^{\mu\rho}g^{\nu\sigma}+g^{\mu\sigma}g^{\nu\rho})$, ${\rm tr}(\gamma^\mu\gamma^\nu\gamma^\rho\gamma^\sigma\gamma^5)=-4i\epsilon^{\mu\nu\rho\sigma}$, where $\epsilon^{0123}=-\epsilon_{0123}=+1$, one finally obtains
\begin{eqnarray}
H_{\rm \eta\eta'}^{\mu\nu}&=&
2\lbrack k^{\eta,\mu}_1k^{\eta',\nu}_2-g^{\mu\nu}k^\eta_1k_2^{\eta'}+k^{\eta,\nu}_1k^{\eta',\mu}_2 \cr
&+&i\epsilon^{\alpha\mu\beta\nu}k^\eta_{1\alpha}k^{\eta'}_{2\beta}\rbrack \ ,
\label{eq:qtensor}
\end{eqnarray}
where 
\beqa
k_1^\eta&=&\left(p_1^\eta-\frac{q}{2}\right)\left(1-p_1^\eta Q_1^\eta\right)+m^2Q_1^\eta,\nonumber\\
k_2^{\eta'}&=&\left(p_2^{\eta'}+\frac{q}{2}\right)\left(1-p_2^{\eta'} Q_2^{\eta'}\right)+m^2Q_2^{\eta'},
\eeqa
with $Q_i^\eta=-\eta q/\sqrt{(p_1^\eta q)^2-m^2 q^2}$.
The spin polarization for $u$ quark is denoted as $\eta'$.

\section{Angular integral: near the onset density of the DCDW}
\label{sec:c}

In Appendix~\ref{sec:c}, we evaluate the angular integral $\tilde A$ [Eq.~(\ref{eq:angleonset3})],  
\begin{equation}
\tilde A=\left(\prod_{i=1}^3 \int d\Omega_i \right)(1-\cos\theta_{23})\int d^3 x e^{i(\bp_1-\bp_2-\bp_3-\bq)\cdot x} \ . 
 \label{eq:angleonsetc}
\end{equation}
$\tilde A$ is separated into two parts:
$\tilde A=\tilde A_1+\tilde A_2$ with
\begin{equation}
\tilde A_1=\left(\prod_{i=1}^3 \int d\Omega_i \right) \int d^3 x {\rm exp}\left[i(\bp_1-\bp_2-\bp_3-\bq)\cdot\bx\right] 
 \label{eq:tildea1}
\end{equation}
and 
\begin{eqnarray}
\tilde A_2&=&-\left(\prod_{i=1}^3 \int d\Omega_i \right) \cos\theta_{23} \cr
&\times&\int d^3 x {\rm exp}\left[i(\bp_1-\bp_2-\bp_3-\bq)\cdot\bx\right] \ . 
 \label{eq:tildea2}
\end{eqnarray}
The $\tilde A_1$ is represented by the use of the spherical Bessel function $j_0(x)$ as 
\begin{eqnarray}
\tilde A_1&=&\int d^3 x e^{- i {\bf q}\cdot{\bf x}}(4\pi)^3 
j_0(|\bp_1|x) j_0(|\bp_2|x) j_0(|\bp_3|x) \cr
&=& (4\pi)^4\int_0^\infty dx x^2  j_0(|{\bf q}|x) j_0(|\bp_1|x) j_0(|\bp_2|x) j_0(|\bp_3|x)\cr
&=&\frac{64\pi^5}{|{\bf p}_1| |{\bf p}_2|  |{\bf q}|}
\label{eq:tildea11}
\end{eqnarray}
for $| |{\bf p}_1| - |{\bf p}_2| |+|{\bf p}_3|< |{\bf q}| < |{\bf p}_1| + |{\bf p}_2|-|{\bf p}_3|$.  
This kinematical condition is met in the case near the onset density, since $|{\bf p}_1|\sim |{\bf p}_2|\sim \mu_{\rm c1}=0.49\Lambda \gg |{\bf p}_3|\sim \mu_e$, and $|{\bf q}|=0.55\Lambda$ (see Table~\ref{para}). 

Next consider the $\tilde A_2$. 
By expanding $\cos\theta_{23}$ in terms of the spherical harmonics, one has
\begin{eqnarray}
&&\tilde A_2=-\int d^3 x e^{- i{\bf q}\cdot{\bf x}}(4\pi)j_0(|{\bf p}_1|x)\int d\Omega_2\int d\Omega_3 \cr
&\times&\frac{4\pi}{3}\sum_{M=-1}^{1} Y_1^{M\ast}(\Omega_2)Y_1^{M}(\Omega_3) e^{-i|{\bf p}_2|x\cos\theta_2} e^{-i|{\bf p}_3|x\cos\theta_3} \cr
&=&(4\pi)^4\int_0^\infty d x x^2
 j_0(|{\bf q}|x) j_0(|\bp_1|x) j_1(|\bp_2|x) j_1(|\bp_3|x)\ , \cr
&& \label{eq:tildea21}
\end{eqnarray}
where $\theta_i$ ($i$=2,3) is the angle between ${\bf p}_i$ and ${\bf x}$.
With $|{\bf p}_3| \sim \mu_e \ll \mu_{\rm c1}$, numerical estimation shows $\tilde A_2 \ll \tilde A_1$, so that we can safely neglect the $\tilde A_2$ in $\tilde A$ as compared with $\tilde A_1$.   

\section{Angular integral: near the termination density of the DCDW}
\label{sec:d}

In Appendix~\ref{sec:d}, we evaluate the angular integrals, $\tilde B_1, \tilde B_2, \tilde B_3$, [Eq.~(\ref{eq:Btilde})].

First we consider the $\tilde B_1$: 
\begin{eqnarray}
\tilde B_1&=&-\left(\prod_{i=1}^3\int d\Omega_i\right)\int d^3 x \exp\left[i({\bf p}_1'-{\bf p}_2'-{\bf p}_3)\cdot{\bf x}\right] \cr
&\times&\frac{|{\bf q}|^2}{\sqrt{\left[({\bf p}_1'+{\bf q}/2)\cdot {\bf q}\right]^2+m^2|{\bf q}|^2}} \ .
\label{ad:Btilde1}
\end{eqnarray}
The angular integration over $\bp_2'$ and $\bp_3$ in Eq.~(\ref{ad:Btilde1}) gives
\begin{eqnarray}
\tilde B_1&=& - 4(2\pi)^2|{\bf q}|^2\int d^3 x j_0(|{\bf p}_2'|x) j_0(|{\bf p}_3|x) \cr
&\times&\int d\Omega_1\frac{e^{i{\bf p}_1'\cdot {\bf x}}}{\sqrt{\left[({\bf p}_1'+{\bf q}/2)\cdot {\bf q}\right]^2+m^2|{\bf q}|^2}} \ .
\label{ad:Btilde12}
\end{eqnarray}
Here the factor $e^{-i\bp'_1\cdot\bx}$ can be expanded in terms of the spherical Bessel functions and the spherical harmonics as 
\beqa
e^{i\bp'_1\cdot\bx}&=&\sum_{L,M}(4\pi) i^Lj_L(|\bp'_1|x)Y_L^{M\ast}(\Omega_1)Y_L^{M}(\Omega_x),
\label{ad:exp}
\eeqa
where we have taken as $\bq // {\hat z}$. 
Thus we can evaluate the remaining angular integrations of $\bp'_1$ and $\bx$.
By the use of the relation, 
$\displaystyle \int d\Omega_x Y_L^M(\Omega_x)=(4\pi)^{1/2}\delta_{L,0}\delta_{M,0}$, one obtains
\begin{eqnarray}
\tilde B_1&=&-64\pi^3 |{\bf q}|^2 \int_0^\infty dx x^2  j_0(|{\bf p}_1'|x) j_0(|{\bf p}_2'|x) j_0(|{\bf p}_3|x) \cr
&\times&\int d\Omega_1\frac{1}{\sqrt{\left(|{\bf p}_1'||{\bf q}|\cos\theta_1+|{\bf q}|^2/2\right)^2+m^2|{\bf q}|^2} } \ .
\label{eq:Bu4}
\end{eqnarray}
In Eq.~(\ref{eq:Bu4}), 
\begin{equation}
\int_0^\infty dx x^2  j_0(|{\bf p}_1'|x) j_0(|{\bf p}_2'|x) j_0(|{\bf p}_3|x)=\frac{\pi}{4|{\bf p}_1'| |{\bf p}_2'| |{\bf p}_3|} 
\label{eq:ssj}
\end{equation}
 for $ ||{\bf p}_1'|-|{\bf p}_2' | |<|{\bf p}_3|< |{\bf p}_1'|+|{\bf p}_2' |$,
and 
\begin{eqnarray}
\int &d\Omega_1&\frac{1}{\sqrt{\left(|{\bf p}_1'||{\bf q}|\cos\theta_1+|{\bf q}|^2/2\right)^2+m^2|{\bf q}|^2} }\cr
&=&\frac{2\pi}{|{\bf p}_1'||{\bf q}|}I(\frac{ |{\bf q}|}{2|{\bf p}_1'|}, \frac{m}{|{\bf p}_1'|}) 
\label{eq:angle1}
\end{eqnarray}
with
\begin{equation}
I(a, b)\equiv\log\left\vert\frac{a+1+\sqrt{(a+1)^2+b^2}}{a-1+\sqrt{(a-1)^2+b^2}}\right\vert \ .
\label{eq:i}
\end{equation}
Substituting Eqs.~(\ref{eq:ssj}), (\ref{eq:angle1}), (\ref{eq:i}) into Eq.~(\ref{eq:Bu4}), one obtains
\begin{equation}
\tilde B_1\simeq -32\pi^5 \frac{|{\bf q}|}{|{\bf p}_1'|^2 |{\bf p}_2'| |{\bf p}_3|}\log\left(\frac{2|{\bf q}|}{m}\right) \ , 
\label{eq:B1tildefinal}
\end{equation}
where we have used $|\bp_1'|\simeq |\bq|/2$ and $m\ll |{\bf p}_1'|$. 

Second, the $\tilde B_2$ is calculated in a way similar to the case of the $\tilde B_1$. The result is
\begin{equation}
\tilde B_2\simeq 32\pi^5 \frac{|{\bf q}|}{|{\bf p}_1'| |{\bf p}_2'|^2 |{\bf p}_3|}\log\left(\frac{2|{\bf q}|}{m}\right) \ . 
\label{eq:B2tildefinal}
\end{equation}

Finally, we consider $\tilde B_3$ :
\begin{eqnarray}
\tilde B_3&=&\left(\prod_{i=1}^3\int d\Omega_i\right)\int d^3 x \exp\left[i({\bf p}_1'-{\bf p}_2'-{\bf p}_3)\cdot{\bf x}\right] \cr
&\times&\frac{2({\bf p}_3\cdot{\bf q})}{\sqrt{\left[({\bf p}_2'-{\bf q}/2)\cdot {\bf q}\right]^2+m^2|{\bf q}|^2}} \ .
\label{ad:Btilde3}
\end{eqnarray}

The angular integration over ${\bf p}_1'$ gives
\begin{eqnarray}
\tilde B_3&=&8\pi\int dx x^2\int d\Omega_x  j_0(|{\bf p}_1'|x) \int d\Omega_2\int d\Omega_3 \cr
&\times&4\pi\sum_{L_2=0}^\infty\sum_{M_2=-L_2}^{L_2}(-i)^{L_2}j_{L_2}(|{\bf p}_2'|x)Y_{L_2}^{M_2}(\Omega_2)Y_{L_2}^{M_2\ast}(\Omega_x) \cr
&\times&4\pi\sum_{L_3=0}^\infty\sum_{M_3=-L_3}^{L_3}(-i)^{L_3}j_{L_3}(|{\bf p}_3|x)Y_{L_3}^{M_3}(\Omega_3)Y_{L_3}^{M_3\ast}(\Omega_x) \cr
&\times&\frac{|{\bf p}_3||{\bf q}|\cos\theta_3}{\sqrt{\left(|{\bf p}_2'||{\bf q}|\cos\theta_2-|{\bf q}|^2/2\right)^2+m^2|{\bf q}|^2} } \ , 
\label{eq:Btilde3}
\end{eqnarray}
where the remaining two exponential factors, \break $\exp(-i{\bf p}_2'\cdot {\bf x})$ and $\exp(-i{\bf p}_3\cdot {\bf x})$ in Eq.~(\ref{ad:Btilde3}), have been expanded in terms of the spherical Bessel functions and the spherical harmonics.
$\displaystyle Y_{L_3}^{M_3\ast}(\Omega_x) $ in Eq.~(\ref{eq:Btilde3}) is rewritten as $\displaystyle Y_{L_3}^{M_3\ast}(\Omega_x)=(-1)^{M_3}Y_{L_3}^{-M_3}(\Omega_x)$. Then,   by the help of the orthonormality of the spherical harmonics, $\displaystyle\int d\Omega_x Y_{L_2}^{M_2\ast}(\Omega_x) Y_{L_3}^{M_3}(\Omega_x)=\delta_{L_2, L_3}\delta_{M_2, M_3} $, Eq.~(\ref{eq:Btilde3}) reads
\begin{eqnarray}
\tilde B_3&=&128\pi^3 |{\bf p}_3||{\bf q}| \sum_{L_2=0}^\infty\sum_{M_2=-L_2}^{L_2}(-1)^{L_2} \cr
&\times&\int_0^\infty dx x^2 j_0(|{\bf p}_1'|x)  j_{L_2}(|{\bf p}_2'|x) j_{L_2}(|{\bf p}_3|x) \cr
&\times&\int d\Omega_2 \frac{Y_{L_2}^{M_2}(\Omega_2)}{\sqrt{\left(|{\bf p}_2'||{\bf q}|\cos\theta_2-|{\bf q}|^2/2\right)^2+m^2|{\bf q}|^2} } \cr
&\times&\int d\Omega_3 (-1)^{M_2}Y_{L_2}^{-M_2}(\Omega_3)\cos\theta_3 \ . 
\label{eq:B35}
\end{eqnarray}
The last integral in Eq.~(\ref{eq:B35}) with respect to the angle of the ${\bf p}_3$ yields $\sqrt{4\pi/3}\delta_{L_2,1}\delta_{M_2,0}$. Thereby Eq.~(\ref{eq:B35}) reads
\begin{equation}
\tilde B_3 = - 128\pi ^3 |{\bf p}_3||{\bf q}| PQ \ , 
\label{eq:B36}
\end{equation}
where
\begin{eqnarray}
P&\equiv& \int_0^\infty dx x^2 j_0(|{\bf p}_1'|x)  j_1(|{\bf p}_2'|x)j_1(|{\bf p}_3|x) \cr
&=&\frac{\pi}{8}\frac{|{\bf p}_2'|^2+|{\bf p}_3|^2-|{\bf p}_1'|^2}{|{\bf p}_1'| |{\bf p}_2'|^2 |{\bf p}_3|^2} \ (< 0) \ ,
\label{eq:P}
\end{eqnarray}
which is valid for $\quad ||{\bf p}_1'|-|{\bf p}_2' | |<|{\bf p}_3|< |{\bf p}_1'|+|{\bf p}_2' |$, 
and 
\begin{eqnarray}
Q&\equiv&\int d\Omega_2\frac{\cos\theta_2}{\sqrt{\left(|{\bf p}_2'||{\bf q}|\cos\theta_2-|{\bf q}|^2/2\right)^2+m^2|{\bf q}|^2} } \cr
&=&\frac{2\pi}{|{\bf p}_2'| |{\bf q}|}J\left(\frac{|{\bf q}|}{2|{\bf p}_2'|}, \frac{m}{|{\bf p}_2'|}\right)  
\label{eq:Q}
\end{eqnarray}
with
\begin{eqnarray}
J(a,b)&\equiv &\left(\sqrt{(1-a)^2+b^2}-\sqrt{(1+a)^2+b^2}\right) \cr
&+& a I(a,b)  \ .
\label{eq:j}
\end{eqnarray}
Since $|\bp_2'|\simeq |\bq|/2$ and $m\ll |{\bf p}_2'|$, we have
\begin{equation}
Q\simeq\frac{\pi}{|{\bf p}_2'|^2}\left\lbrack\log\left(\frac{2|{\bf q}|}{m}\right)-2\right\rbrack \ .
\label{eq:Q2}
\end{equation}
Substituting Eqs.~(\ref{eq:P}) and (\ref{eq:Q2}) into Eq.~(\ref{eq:B36}), one finally obtains
 \begin{eqnarray}
\tilde B_3&\simeq& 16\pi^5|{\bf q}|\frac{|{\bf p}_1'|^2-|{\bf p}_2'|^2-|{\bf p}_3|^2}{|{\bf p}_1'| |{\bf p}_2'|^4|{\bf p}_3|} \cr
&\times&\left[\log\left(\frac{2|{\bf q}|}{m}\right)-2\right] \ .
\label{eq:B3final}
\end{eqnarray}

\end{document}